# 車聯網安全憑證管理系統北美標準 IEEE 1609.2.1 和歐盟標準 ETSI TS 102 941 效能比較

Abel C. H. Chen[*]

Information & Communications Security Laboratory,
Chunghwa Telecom Laboratories

**摘要**

本研究探討車聯網(Vehicular-to-Everything, V2X)通訊中安全憑證管理系統(Security Credential Management Systems, SCMS)的兩大主要標準(北美 IEEE 標準和歐洲 ETSI 標準)之間的效能和結構差異。本研究著重在比較兩者的公鑰基礎設施(Public Key Infrastructure, PKI)架構、憑證申請流程,以及流程中請求/回應訊息的格式。此研究還包含對每個標準中訊息長度和安全特性的理論分析。此外,通過對兩種系統的實現,評估安全憑證管理系統的效率。基於研究結果,本研究提出相關建議,以作為未來部署安全憑證管理系統的參考。

*關鍵詞:車聯網、安全憑證管理系統、IEEE 1609.2.1、ETSI TS 102 941*

# Performance Comparison of Security Credential Management Systems for V2X: North American Standard IEEE 1609.2.1 and European Standard ETSI TS 102 941

**Abstract**

This study examines the performance and structural differences between the two primary standards for Security Credential Management Systems (SCMS) in Vehicular-to-Everything (V2X) communication: the North American IEEE standards and the European ETSI standards. It focuses on comparing their respective Public Key Infrastructure (PKI) architectures, certificate application workflows, and the formats of request/response messages used during certificate applications. The research includes a theoretical analysis of message length and security features inherent in each standard. Additionally, practical implementations of both systems are conducted to evaluate their efficiency in certificate management processes. Based on the findings, the study provides recommendations to guide future development of SCMS.

[*] Corresponding author E-mail: chchen.scholar@gmail.com

車聯網安全憑證管理系統北美標準 IEEE 1609.2.1 和歐盟標準 ETSI TS 102 941 效能比較

# 1.前言

為了實現車載設備(On-Board Units, OBUs)與路側設備(Road-Side Units, RSUs)之間的安全通訊，美國電氣電子工程師學會(Institute of Electrical and Electronics Engineers, IEEE)和歐洲電信標準協會(European Telecommunications Standards Institute, ETSI)逐步制定了相關標準(Li, Wang, Sharma, & Gope, 2023; Yoshizawa & Preneel, 2022)。在 IEEE 標準中，憑證格式由 IEEE 1609.2 規範(Intelligent Transportation Systems Committee, 2023)，而憑證申請流程(包含註冊憑證(Enrollment Certificate)、授權憑證(Authorization Certificate))、請求/回應訊息格式則由 IEEE 1609.2.1 規範(Intelligent Transportation Systems Committee, 2022)。此外，IEEE 標準已被包括北美多個國家/地區採用為車聯網(Vehicular-to-Everything, V2X)通訊的標準格式。另一方面，在 ETSI 標準中，憑證格式由 ETSI TS 103 097 規定(European Telecommunications Standards Institute, 2021)，其憑證格式與 IEEE 1609.2 中的憑證格式非常相似。ETSI TS 102 941 則規範了憑證申請流程(包含註冊憑證(Enrolment Credential)和授權票證(Authorization Ticket))、請求/回應訊息格式(European Telecommunications Standards Institute, 2022)。ETSI 標準已被歐洲多個國家/地區採用為車聯網通訊的標準格式。不同標準在不同國家/地區有各自的支持者，當在某特定國家/地區部署車聯網通訊設備時，需遵守當地法律。因此，了解不同標準之間的差異是一個重要議題。

為實現不同標準之間的憑證相容性，IEEE 1609.2 與 ETSI TS 103 097 規定的憑證格式相似，其欄位資訊一致，但在數值定義上存在細微差異。有鑑於此，本研究聚焦於比較 IEEE 1609.2.1 與 ETSI TS 102 941 標準中的憑證申請流程與請求/回應訊息格式。深入探討 IEEE 的註冊憑證請求/回應與 ETSI 的註冊憑證請求/回應之間的差異，並比較 IEEE 的授權憑證請求/回應與 ETSI 的授權票證請求/回應。此外，本研究還將實作 IEEE 1609.2、IEEE 1609.2.1、ETSI TS 103 097、ETSI TS 102 941 等標準，並在車載設備上進行測試，對不同標準的執行效率進行比較與討論。

本文結構分為五節。第二節介紹 IEEE 標準的安全憑證管理系統(Security Credential Management Systems, SCMS)的架構(Brecht, B. et al., 2018)，以及其憑證申請流程的步驟。第三節說明 ETSI 標準的合作式智慧運輸系統(Cooperative Intelligent Transport Systems, C-ITS)安全憑證管理系統(C-ITS Security Credential Management Systems, CCMS)的架構(Lonc, Aubry, Bakhti, Christofi, & Mehrez, 2023)，其憑證申請流程的步驟。第四節展示 IEEE 安全憑證管理系統與 ETSI 合

作式智慧運輸系統安全憑證管理系統的實作與效能評估。最後,第五節討論本研究的貢獻及未來工作。

## 2.IEEE 安全憑證管理系統(SCMS)

本節描述 IEEE 1609.2.1 標準中規範的安全憑證管理系統。第 2.1 節介紹安全憑證管理系統架構,並且描述整個憑證申請流程。第 2.2 節詳述註冊憑證請求與回應的具體內容,而第 2.3 節則提供了授權憑證請求與回應的詳細說明。

### 2.1 安全憑證管理系統架構和憑證申請流程

在安全憑證管理系統中,根憑證中心(Root Certificate Authority, RCA)具有為自身及中繼憑證中心(Intermediate Certificate Authority, ICA)簽發憑證的能力。此外,ICA 有權為註冊憑證中心(Enrollment Certificate Authority, ECA)、授權憑證中心(Authorization Certificate Authority, ACA)、以及登錄中心(Registration Authority, RA)簽發憑證。終端設備(End Entity, EE)的註冊憑證由註冊憑證中心簽發,而終端設備的授權憑證則由授權憑證中心簽發。所有在安全憑證管理系統內運作的設備均依賴於由根憑證中心起始的憑證鏈(Certificate Chain)來建立信任(Intelligent Transportation Systems Committee, 2022),安全憑證管理系統架構如圖 1 所示。

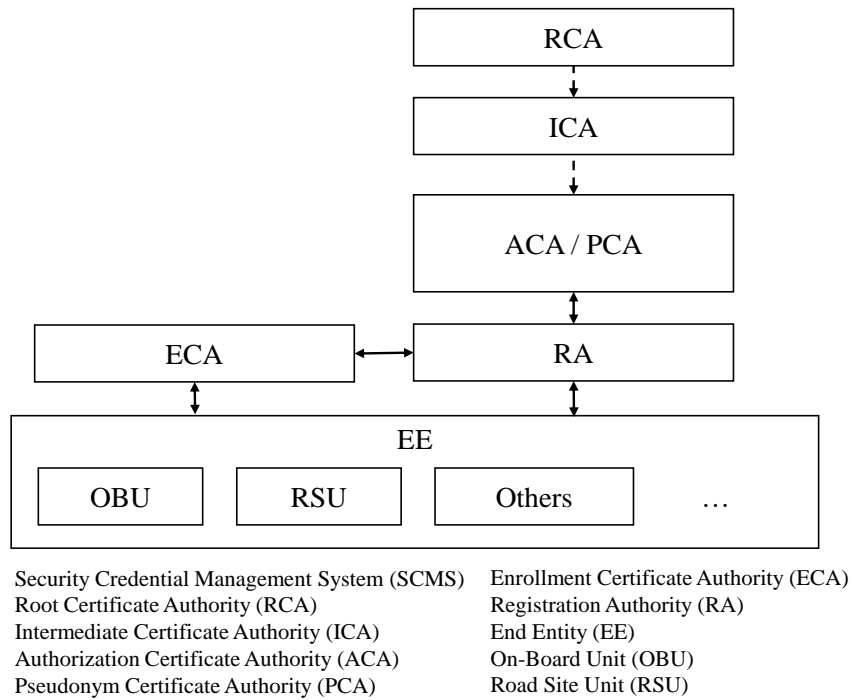

Security Credential Management System (SCMS)　　Enrollment Certificate Authority (ECA)
Root Certificate Authority (RCA)　　Registration Authority (RA)
Intermediate Certificate Authority (ICA)　　End Entity (EE)
Authorization Certificate Authority (ACA)　　On-Board Unit (OBU)
Pseudonym Certificate Authority (PCA)　　Road Site Unit (RSU)

**圖 1　安全憑證管理系統(SCMS)架構**

圖 2 展示安全憑證管理系統內憑證申請流程的詳細流程圖(Intelligent Transportation Systems Committee, 2022)。本研究著重於終端設備的操作步驟，因此將評估包括 EeEcaCertRequestSpdu、EcaEeCertResponseSpdu、EeRaCertRequestSpdu、RaEeCertInfoSpdu 和 AcaResponse 在內的安全協定資料單元(Secure Protocol Data Units, SPDUs)。

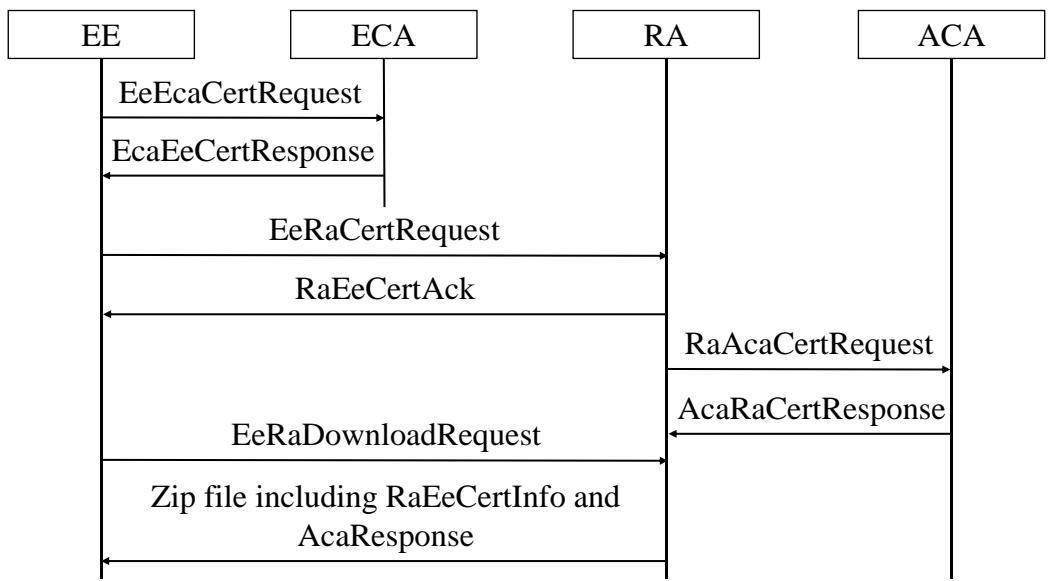

**圖 2　安全憑證管理系統憑證(SCMS)申請流程**

為了申請註冊憑證，終端設備需發送 EeEcaCertRequestSpdu 安全協定資料單元作為發起請求。當註冊憑證中心收到該請求後，會對其進行驗證，若驗證成功，則會在 EcaEeCertResponseSpd 安全協定資料單元中簽發終端設備的註冊憑證(Intelligent Transportation Systems Committee, 2022)。EeEcaCertRequestSpdu 和 EcaEeCertResponseSpdu 的格式詳見第 2.2 節。

在申請授權憑證時(Barreto, Simplicio, Ricardini & Patil, 2021)，終端設備首先發送 EeRaCertRequestSpd 安全協定資料單元作為啟動請求。登錄中心在收到該請求後，會對其進行驗證，並通過 RaEeCertAck 回應，告知驗證結果以及授權憑證的下載時間。若 EeRaCertRequestSpdu 的驗證成功，登錄中心會向授權憑證中心發送 RaAcaCertRequest，並接收包含 AcaResponse 的 AcaRaCertResponse。隨後，終端設備會發送 EeRaDownloadRequest 以下載其授權憑證，RA 則回應一個包含 RaEeCertInfoSpdu 和 AcaResponse 的 zip 壓縮檔，其中包含終端設備的授權憑證(Intelligent Transportation Systems Committee, 2022)。EeRaCertRequestSpdu、RaEeCertInfoSpdu 和 AcaResponse 的格式詳見第 2.3 節。

車聯網安全憑證管理系統北美標準 IEEE 1609.2.1 和歐盟標準 ETSI TS 102 941 效能比較

## 2.2 註冊憑證請求和回應

　　EeEcaCertRequestSpdu 的結構包含 Ieee1609Dot2Data-SignedCertRequest，其中包含 SignedCertificateRequest，並具備如圖 3 所示的簽章資料(Signed Data)結構。該簽章資料結構由簽章者(signer)、簽章(signature)、以及被簽章資料(To-Be-Signed-Data, tbsData)組成。在 tbsData 中，EeEcaCertRequest 包括應用權限(appPermissions)、公鑰、以及其他相關資訊(Intelligent Transportation Systems Committee, 2022)。

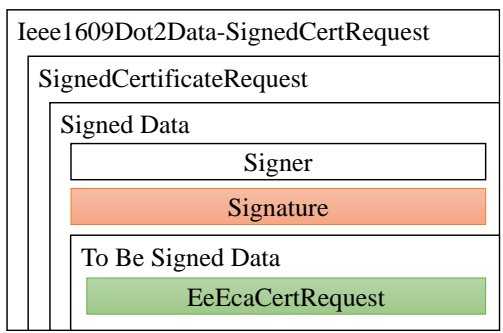

**圖 3　EeEcaCertRequestSpdu 的結構**

　　EcaEeCertResponseSpdu 的結構由 Ieee1609Dot2Data-Signed 組成，封裝 Ieee1609Dot2Content，並具備如圖 4 所示的簽章資料(Signed Data)結構。該簽章資料結構包括簽章者(signer)、簽章(signature)和被簽章資料(tbsData)。在 tbsData 中包含如請求雜湊值(requestHash)、註冊憑證中心憑證鏈(ecaCertChain)、以及註冊憑證等 (Intelligent Transportation Systems Committee, 2022)。EcaEeCertResponseSpdu 中的 ecaCertChain 包含來自根憑證中心、中繼憑證中心、註冊憑證中心的憑證，因此導致 EcaEeCertResponseSpdu 的長度較長，通常超過四個憑證。

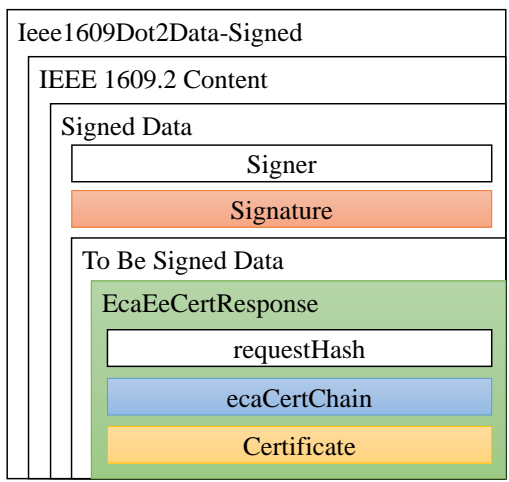

**圖 4　EcaEeCertResponseSpdu 的結構**

## 2.3 授權憑證請求和回應

EeRaCertRequestSpdu 的結構由 Ieee1609Dot2Data-SignedEncryptedCertRequest 組成，其中包含接收者(recipient)和 Ieee1609Dot2Data-SignedCertRequest。後者包含 SignedCertificateRequest，並具備如圖 5 所示的簽章資料(Signed Data)結構。接收者內容是登錄中心公鑰的雜湊值，登錄中心使用其公鑰對 EeRaCertRequestSpdu 進行解密。該簽章資料包括簽章者(signer)、簽章(signature)、以及被簽章資料(tbsData)，其中 EeRaCertRequest 包括應用權限(appPermissions)、公鑰、及其他相關資訊(Intelligent Transportation Systems Committee, 2022)。

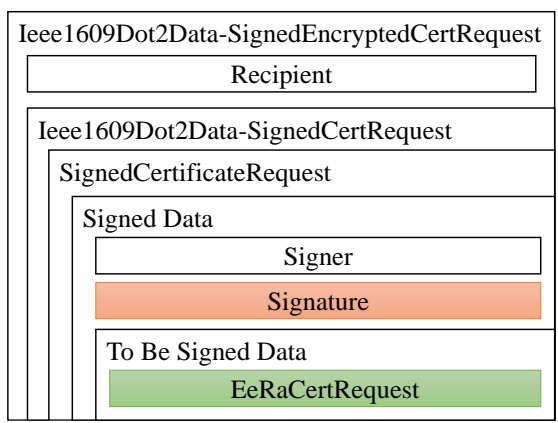

**圖 5　EeRaCertRequestSpdu 的結構**

為了下載授權憑證，登錄中心會向終端設備發送包含 RaEeCertInfoSpdu 和多個 AcaResponse 的 zip 壓縮檔。在 RaEeCertInfoSpdu 中，使用 Ieee1609Dot2Data-Unsecured 格式，其中包含帶有 RaEeCertInfo 的 Ieee1609Dot2Content。RaEeCertInfo 包括請求雜湊值(requestHash)、產製時間(generationTime)、當前時間週期(currentI)、以及下一次下載時間(nextDlTime)等，如圖 6 所示(Intelligent Transportation Systems Committee, 2022)。終端設備可利用 currentI 的值產製繭私鑰(cocoon private key)，並參考 nextDlTime 以便未來下載更新的授權憑證。

在下載授權憑證時，zip 壓縮檔可能包含多個 AcaResponse。每個 AcaResponse 遵循 Ieee1609Dot2Data-EncryptedSigned 格式。終端設備可以使用其繭私鑰對每個 AcaResponse 進行解密，從而取得簽章資料。在每個簽章資料中，包含簽章者(signer)、簽章(signature)和被簽章資料(tbsData)，其中包括 AcaEeCertResponse。AcaEeCertResponse 包括產製時間(generationTime)、私鑰資訊(privateKeyInfo)、以及授權憑證等，如圖 7 所示(Intelligent Transportation Systems Committee,

2022)。此外，privateKeyInfo 可以用於將繭私鑰擴展為蝴蝶私鑰(butterfly private key)。

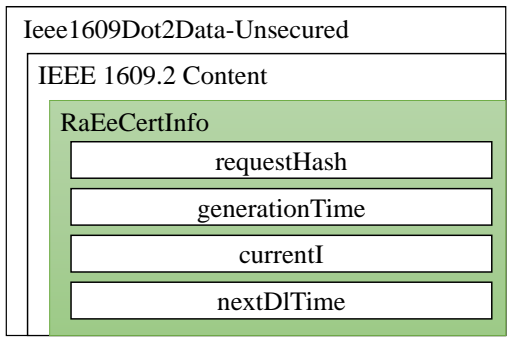

圖 6　　**RaEeCertInfoSpdu** 的結構

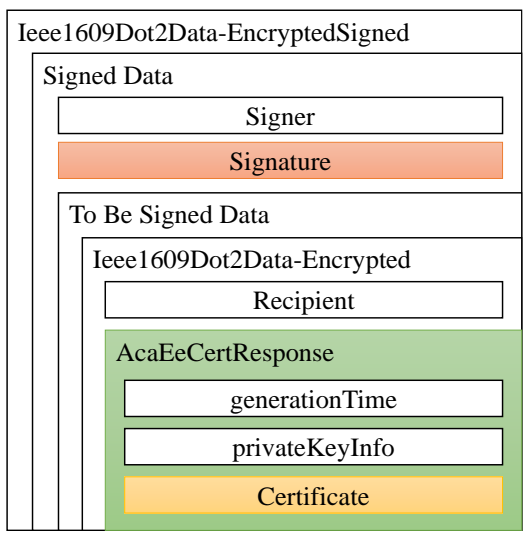

圖 7　　**AcaResponse** 的結構

# 3.ETSI 合作式智慧運輸系統安全憑證管理系統(CCMS)

本節描述 ETSI TS 102 941 標準中定義的合作式智慧運輸系統安全憑證管理系統。第 3.1 節描述合作式智慧運輸系統安全憑證管理系統架構，並闡述整個憑證申請流程。第 3.2 節詳述註冊憑證請求與回應的細節，而第 3.3 節則提供授權票證請求與回應的相關資訊。

## 3.1 合作式智慧運輸系統安全憑證管理系統架構

ETSI 合作式智慧運輸系統安全憑證管理系統架構相比於 IEEE 安全憑證管理系統具有更扁平化的結構，特別是不包含中繼憑證中心。在此架構中，根憑證中心直接為註冊中心(Enrolment Authority, EA)和授權中心(Authorization

Authority, AA)簽發憑證(如圖 8 所示)(European Telecommunications Standards Institute, 2022; Lone, Verma, & Sharma, 2024),因此不需要登錄中心角色。在這種架構下,合作式智慧運輸系統安全憑證管理系統中的註冊中心功能類似於安全憑證管理系統中的註冊憑證中心,而合作式智慧運輸系統安全憑證管理系統中的授權中心功能則類似於安全憑證管理系統中的授權憑證中心。此外,合作式智慧運輸系統安全憑證管理系統中的註冊中心還承擔了安全憑證管理系統中登錄中心的角色。車載設備和路側設備被歸類為智慧運輸系統站點(Intelligent Transport System Station, ITS-S);註冊中心有能力為這些為智慧運輸系統站點簽發註冊憑證(類似於 IEEE 1609.2.1 中的註冊憑證),而授權中心則可以為其簽發授權票證(類似於 IEEE 1609.2.1 中的授權憑證)。

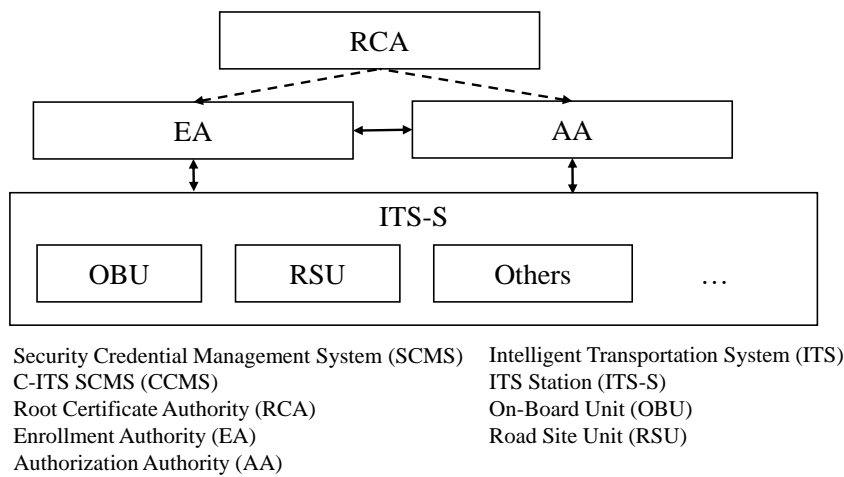

Security Credential Management System (SCMS)　　Intelligent Transportation System (ITS)
C-ITS SCMS (CCMS)　　ITS Station (ITS-S)
Root Certificate Authority (RCA)　　On-Board Unit (OBU)
Enrollment Authority (EA)　　Road Site Unit (RSU)
Authorization Authority (AA)

**圖 8　合作式智慧運輸系統安全憑證管理系統(CCMS)架構**

圖 9 展示合作式智慧運輸系統安全憑證管理系統內憑證申請流程的詳細流程圖(European Telecommunications Standards Institute, 2022)。本研究著重於智慧運輸系統站點的操作步驟,因此將評估包括 EnrolmentRequest、EnrolmentResponse、AuthorizationRequest 和 AuthorizationResponse 等訊息。

為請求註冊憑證,智慧運輸系統站點首先通過發送 EnrolmentRequest 的訊息來請求註冊憑證。註冊中心收到此請求後進行驗證,如果驗證成功,則會在 EnrolmentResponse 的訊息中簽發智慧運輸系統站點的註冊憑證(European Telecommunications Standards Institute, 2022)。EnrolmentRequest 和 EnrolmentResponse 訊息的格式詳見第 3.2 節。

在請求授權票證的過程中,智慧運輸系統站點發送 AuthorizationRequest 的訊息並附帶一個 EcSignature。當授權中心收到此請求後進行驗證,隨後將帶有 EcSignature 的 AuthorizationValidationRequest 訊息發送至註冊中心進行驗證。註

冊中心會回應 AuthorizationValidationResponse。如果驗證成功，授權中心會簽發授權票證，並將包含授權票證的 AuthorizationResponse 訊息返回給智慧運輸系統站點。AuthorizationRequest 和 AuthorizationResponse 的格式詳第 3.3 節。

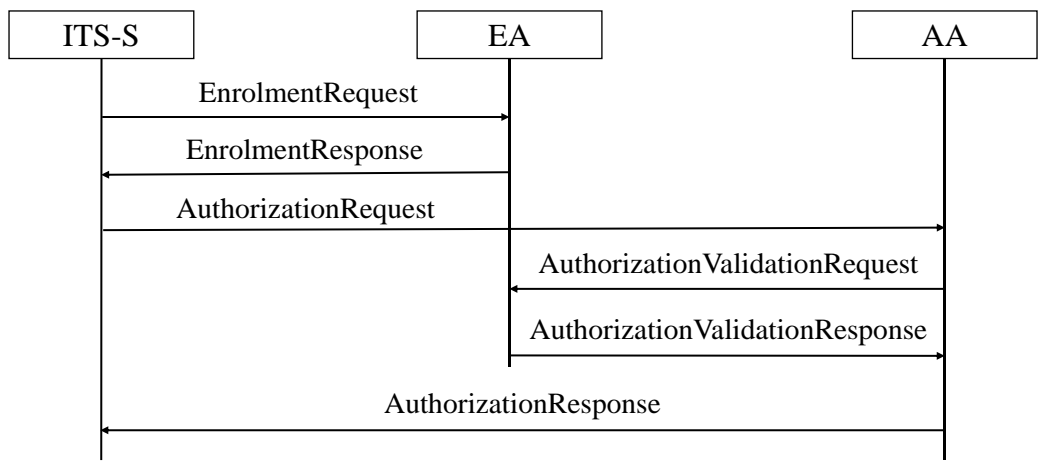

**圖 9　合作式智慧運輸系統安全憑證管理系統(CCMS)憑證申請流程**

## 3.2 註冊憑證請求和回應

每個憑證申請訊息的格式遵循 ETSI TS 103 097 標準，具體為 EtsiTs103097Data-Encrypted，要求使用接收方的公鑰或預共享金鑰(pre-shared key)進行加密。

解密後的 EnrolmentRequest 結構包含接收者內容(註冊中心公鑰的雜湊值)、以及 EtsiTs103097Data-Signed 結構，該結構內包含簽章資料結構(如圖 10 所示)。此簽章資料包含簽章者(signer)、由智慧運輸系統站點的權威私鑰(canonical private key)產製的簽章(signature)和被簽章資料(tbsData)。在 tbsData 中，嵌套了另一個 EtsiTs103097Data-Signed 結構，為了區別，稱為內部簽章資料(inner Signed Data)。內部簽章資料包含簽章者(signer)、由智慧運輸系統站點的註冊私鑰(enrolment private key)產製的簽章(signature)和被簽章資料(tbsData)。內部簽章資料中的 tbsData 包含 InnerEcRequest，其中包括應用權限(appPermissions)、公鑰、以及其他相關資訊(European Telecommunications Standards Institute, 2022)。

在成功驗證請求後，註冊中心根據 InnerEcRequest 產製 EnrolmentResponse。該 EnrolmentResponse 使用在產製 EnrolmentRequest 過程中通過橢圓曲線整合加密方案(Elliptic Curve Integrated Encryption Scheme, ECIES)獲得的預共享金鑰進行加密。

解密後的 EnrolmentResponse 結構包含接收者內容(預共享金鑰的雜湊值)、

以及 EtsiTs103097Data-Signed 結構,該結構內包含簽章資料結構(如圖 11 所示)。此簽章資料包括簽章者(signer)、由註冊中心的私鑰產製的簽章(signature)和被簽章資料(tbsData)。在 tbsData 中,包含一個 InnerEcResponse,其中包括如 requestHash、responseCode 和註冊憑證等(European Telecommunications Standards Institute, 2022)。智慧運輸系統站點從 InnerEcResponse 中獲得其註冊憑證。

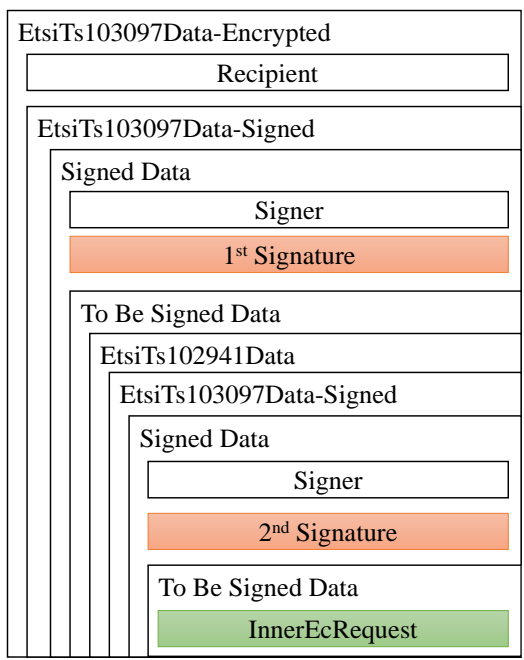

圖 10　EnrolmentRequest 訊息的結構

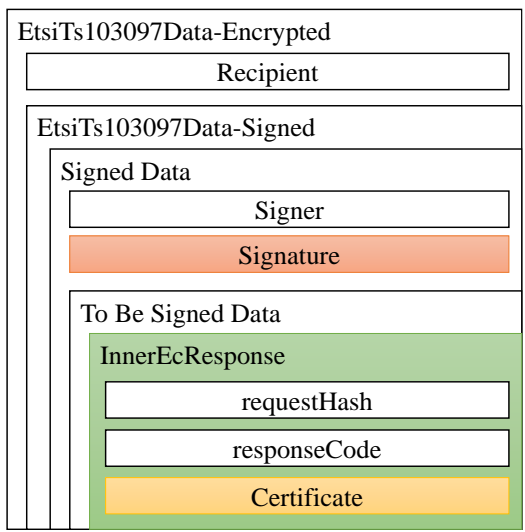

圖 11　EnrolmentResponse 訊息的結構

## 3.3 授權票證請求和回應

在申請授權票證的過程中,解密後的 AuthorizationRequest 結構包含接收者

車聯網安全憑證管理系統北美標準 IEEE 1609.2.1 和歐盟標準 ETSI TS 102 941 效能比較

內容(授權中心公鑰的雜湊值)，並包含 EtsiTs103097Data-Signed 結構，該結構內包含簽章資料(如圖 12 所示)。此簽章資料包括簽章者(signer)、由智慧運輸系統站點的新的授權私鑰(authorization private key)產製的簽章(signature)和被簽章資料(tbsData)。在 tbsData 中，封裝了 InnerAtRequest，其中包含如 SharedAtRequest 和 EcSignature 等(European Telecommunications Standards Institute, 2022)。

為了進行驗證，EcSignature (EtsiTs103097Data-Encrypted 結構)由智慧運輸系統站點使用註冊中心的公鑰加密。EcSignature 包含接收者內容(註冊中心公鑰的雜湊值)，並包含 EtsiTs103097Data-SignedExternalPayload。此 EtsiTs103097Data-SignedExternalPayload 內容包含基於智慧運輸系統站點註冊憑證摘要的簽章者(signer)、由智慧運輸系統站點的註冊私鑰產製的簽章(signature)和被簽章資料(tbsData)(European Telecommunications Standards Institute, 2022)。

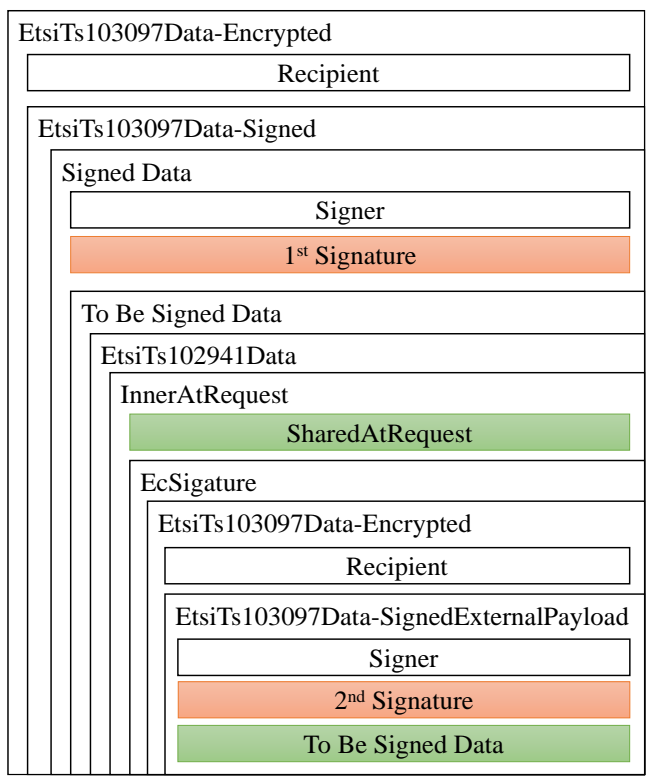

圖 12 **AuthorizationRequest** 訊息的結構

在成功驗證請求後，授權中心根據 SharedAtRequest 產製 AuthorizationResponse。該 AuthorizationResponse 使用在產製 AuthorizationRequest 過程中通過橢圓曲線整合加密方案獲得的預共享密鑰進行加密。

解密後的 AuthorizationResponse 訊息結構包含接收者內容(授權中心公鑰的雜湊值)，以及 EtsiTs103097Data-Signed 結構，該結構內包含簽章資料結構(如圖 13所示)。此簽章資料包括簽章者(signer)、由授權中心的私鑰產製的簽章(signature)

車聯網安全憑證管理系統北美標準 IEEE 1609.2.1 和歐盟標準 ETSI TS 102 941 效能比較

和被簽章資料(tbsData)。在 tbsData 中，包含 AuthorizationResponse，其中包括含 requestHash、responseCode 和授權票證等(European Telecommunications Standards Institute, 2022)。智慧運輸系統站點從 AuthorizationResponse 中取得其授權票證。

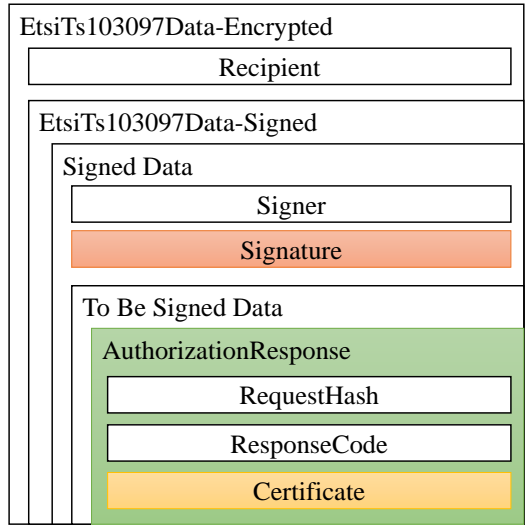

**圖 13　AuthorizationResponse 訊息的結構**

## 4.實驗比較與討論

為比較 IEEE 安全憑證管理系統和 ETSI 合作式智慧運輸系統安全憑證管理系統，本研究實作和評估這兩個系統的效能。第 4.1 節概述實驗環境，第 4.2 節呈現有關訊息長度的比較結果，而第 4.3 節討論計算時間的比較。

### 4.1 實驗環境

在實驗中，作者作為 IEEE 安全憑證管理系統和 ETSI 合作式智慧運輸系統安全憑證管理系統的憑證提供者，並提供公開金鑰基礎建設(Public Key Infrastructure, PKI)服務和憑證。終端設備和智慧運輸系統站點可以根據圖 2 和圖 9 中的流程啟動憑證申請。此外，本研究採用了公信電子的車載設備來實現。終端設備和智慧運輸系統站點的功能。該設備採用作業系統 Android 10 並將權威私鑰整合到終端設備和智慧運輸系統站點中以產製初始的安全協定資料單元或訊息。

### 4.2 訊息長度比較

表 1 顯示基於 IEEE 標準的安全協定資料單元長度和基於 ETSI 標準的憑證申請請求和回應的訊息長度。在 ETSI 合作式智慧運輸系統安全憑證管理系統

中，EnrolmentRequest 和 AuthorizationRequest 訊息包含兩個簽章，導 2 玫它們的長度相比 IEEE 安全憑證管理系統中的 EeEcaCertRequest 和 EeRaCertRequest 來得更長。然而，在 IEEE 安全憑證管理系統中，EcaEeCertResponse 包含 ecaCertChain，因此它的長度比 ETSI 合作式智慧運輸系統安全憑證管理系統中的 EnrolmentResponse 長。RaEeCertInfo 和 AcaResponse 的總長度與 AuthorizationResponse 的長度相似。

表 1　訊息長度比較(單位：Byte)

| IEEE | | ETSI | |
|---|---|---|---|
| EeEcaCertRequestSpdu | 151 | EnrolmentRequest | 424 |
| EcaEeCertResponseSpdu | 957 | EnrolmentResponse | 562 |
| EeRaCertRequestSpdu | 369 | AuthorizationRequest | 590 |
| RaEeCertInfoSpdu | 26 | AuthorizationResponse | 627 |
| AcaResponse | 464 | | |

## 4.3 計算時間比較

表 2 展示在憑證申請流程中，由終端設備和智慧運輸系統站點傳輸的基於 IEEE 標準的安全協定資料單元和基於 ETSI 標準的訊息的產製時間或驗證時間。由於在 ETSI 合作式智慧運輸系統安全憑證管理系統中，憑證申請的每條訊息都需要進行加密和解密，因此 ETSI 標準可能需要更多的計算時間。此外，ETSI 合作式智慧運輸系統安全憑證管理系統中 EnrolmentRequest 和 AuthorizationRequest 的訊息需要產製兩個簽章，而 IEEE 安全憑證管理系統中 EeEcaCertRequest 和 EeRaCertRequest 的安全協定資料單元僅產製一個簽章。因此，ETSI 合作式智慧運輸系統安全憑證管理系統中 EnrolmentRequest 和 AuthorizationRequest 訊息的產製時間比 IEEE 安全憑證管理系統中 EeEcaCertRequest 和 EeRaCertRequest 安全協定資料單元的產製時間更長。

然而，在 IEEE 安全憑證管理系統中，EcaEeCertResponse 包含超過三個憑證的 ecaCertChain，因此需要額外的時間來驗證憑證鏈。結果導致 IEEE 安全憑證

管理系統中 EcaEeCertResponse 的計算時間比 ETSI 合作式智慧運輸系統安全憑證管理系統中 EnrolmentResponse 計算時間更長。此外，在 IEEE 安全憑證管理系統中，解壓包含 RaEeCertInfo 和 AcaResponse 的 Zip 壓縮檔可能還需要額外的計算時間。

表 **2** 計算時間長度比較(單位：毫秒)

| IEEE | | ETSI | |
|---|---|---|---|
| EeEcaCertRequestSpdu | 46 | EnrolmentRequest | 244 |
| EcaEeCertResponseSpdu | 396 | EnrolmentResponse | 171 |
| EeRaCertRequestSpdu | 161 | AuthorizationRequest | 345 |
| RaEeCertInfoSpdu and AcaResponse | 348 | AuthorizationResponse | 171 |

## **5.**結論與未來研究

目前車聯網安全憑證管理系統主要採用兩種標準，即 IEEE 1609.2.1 和 ETSI TS 102 941，本研究主要探討這兩種標準之間的差異。研究總結並比較 IEEE 安全憑證管理系統和 ETSI 合作式智慧運輸系統安全憑證管理系統的架構和憑證申請流程。此外，研究還對憑證申請流程程中每個安全協定資料單元和訊息的長度及計算時間進行比較。

在未來研究中，可以利用兩種標準的優勢並將其整合可能會更具優勢。此外，可以使用共同簽章演算法將兩個簽章合併為一個簽章，以減少訊息長度。

## 致謝

車聯網安全憑證管理系統北美標準 IEEE 1609.2.1 和歐盟標準 ETSI TS 102 941 效能比較

# 參考文獻